\documentclass[12pt,english]{iopart}
\pdfoutput=1
\usepackage[T1]{fontenc}
\usepackage[latin9]{inputenc}
\usepackage{geometry}
\usepackage{color}
\usepackage{array}
\usepackage{float}
\usepackage{calc}
\usepackage{amssymb}
\usepackage{mathdots}
\usepackage{graphicx}
\usepackage{setspace}
\makeatletter

\providecommand{\tabularnewline}{\\}
\newcommand{\lyxdot}{.}

\usepackage{iopams}
\usepackage{setstack}

\usepackage{indentfirst}

\makeatother

\usepackage{babel}
\begin{document}
\title[Plasmonic response of a nanorod in the vicinity of a metallic surface ]{Plasmonic response of a nanorod in the vicinity of a metallic surface: local approach with analytical solution}
\author{I. M. Vasilevskiy$^{1}$ and N. M. R. Peres$^{1,2}$}
\address{$^{1}$University of Minho, Department of Physics and Center of Physics
and Quantalab, 4710-057 Braga, Portugal}
\address{$^{2}$International Iberian Nanotechnology Laboratory (INL), Avenida
Mestre José Veiga, 4715-330 Braga, Portugal}
\begin{abstract}
{\normalsize{}In this paper we present an analytical solution for
the eigenmodes and corresponding electric field}\textcolor{black}{\normalsize{}s}{\normalsize{}
of a composite system made of a nanorod in the vicinity of a plasmonic
semi-infinite metallic system. To be specific, we choose Silver as
the material for both the nanorod and the semi-infinite metal. The
system is composed of two sub-systems with different symmetries: the
rod has }\textcolor{black}{\normalsize{}axial}{\normalsize{} symmetry,
while the interface has a rectangular one. Using a boundary integral
method,}\textcolor{black}{\normalsize{} proposed by}{\normalsize{}
Eyges, we are able to compute analytically the integrals that sew
together the two systems. In the end, the problem is reduced to}\textcolor{black}{\normalsize{}
a o}{\normalsize{}ne of linear algebra, where all the terms in the
system are known analytically. For large distances between the rod
and the planar surface, only a few of those integrals are needed and
a full analytical solution can be obtained. Our results are important
to benchmark other numerical approaches and represent a starting point
in the discussion of systems composed of nanorods and two-dimensional
materials.}{\normalsize\par}
\end{abstract}
\maketitle

\section{Introduction}

Nanophotonics, involving plasmonic components, is nowadays at the
heart of many different techonologies, such as gas sensors, efficient
solar cells, highly sensitive photodetectors and photonic devices
\cite{Monticone2017}. The ability to confine electromagnetic radiation,
spanning the spectral range from the THz to the UV, into tiny geometric
spaces is one of the victories of modern nanophotonics. From the THz
to the mid-IR, graphene can work as a plasmonic material \cite{Stockman2018,Klimov2013,Pelton2013,Maier2007},
showing outstanding confinement of electromagnetic radiation down
to one atom thickness. On the other hand, more conventional materials
such as Silver, Gold, and Copper cover the spectral range from the
near-IR to the UV, allowing strong electromagnetic confinement in
a region not accessible to graphene plasmons. The large degree of
electromagnetic field enhancement is achieved in gaps between two
metallic structures, spheres or cubes, and between spheres, infinite/semi-infinite
rods \cite{Sun2014}, cubes and a metallic/conductive flat interface.

\textcolor{black}{Several mechanisms concerning plasmonics can be
used to efficiently attain light trapping. In the near-field regime,
the confinement of electromagnetic waves due to metallic nanostructures
leads to localized surface plasmons (LSP). The same nanostructures
can also act as launchers of propagation surface plasmon polaritons
(SPP) when they are near a semi-infinite planar metallic interface.
In plasmon cavity mode there is generation of standing waves and the
originated field is independent of the incident light polarization
\cite{Moreau2012}. Another approach to enhance absorption is to focus
the incident light beam to the desired location. Metasurfaces are
thin flat structures able to change phase, amplitude and polarization
of the incident waves \cite{Kildishev2013}, favoring the formation
of incident wavefronts with specific characteristics. These two-dimensional
metamaterials are used for engineering of diverse flat optical devices,
such as flat lenses, mirrors, absorbers and anti-reflection layers,
namely for the production of solar cells \cite{Olaimat2021}. }

The position of the plasmonic resonance depends on many variables:
type of plasmonic material, size and shape of the nanoparticle/rod
and dielectric environment. All these possibilities give us a certain
degree of freedom to tailor the position of the plasmonic resonance
at will. The calculation of the plasmonic resonance is straightforward
in simple geometries, on condition that retardation effects are ignored.
If the system possesses a high degree of symmetry, such as isolated
spheres, rods, toroids, and even combinations of these in certain
cases, the analytical calculation of the eigenfrequencies is possible.
On the other hand, the integration of two plasmonic systems of different
symmetries poses considerable analytical difficulties because it is
not an easy task to match the two disparate symmetries together. 

Methods for solving plasmonic problems are abundant, most of them
relying on the numerical solution of Maxwell's equations both in time
and frequency domains. This approach is nowadays integrated into several
commercial software packages. However, it is still important to find
analytical solutions to the class of problems we are considering in
this paper, allowing to benchmark the numerical methods. Also, analytical
solutions frequently put in evidence physical details that may be
buried in numerical simulations. In addition to the calculation of
the eigenmodes and the fields, other relevant properties of the plasmonic
particles and rods, such as their polarizability, are important to
extract analytically \cite{Jung_2012}. The extension of the analytical
approach to plasmonic systems composed of metallic nanoparticles/nanorods
and two-dimensional materials are also of interest.

As already noted, theoretical studies of polarizability and field
enhancement in cylindrical configurations can be obtained from both
analytical \cite{Schieber1999,Lekner2013} and numerical \cite{Venermo2005}
methods, being of substantial interest due to diverse physics and
engineering applications, such as in the field of plasmonics. The
most common recipe for electrostatic problems consists in solving
Poisson's equation in differential form. There is, however, a long
history of integral method approach to electrostatics, dating back
to the first half of the last century \cite{Phillips1934,Kellogg1967}.
Such methods are very convenient for numerical implementation, as
the problem can be transformed into the solution of an exercise in
linear algebra. Yet, following this approach there are not many known
analytical solutions. In this paper, we present one such solution,
where we match together two systems with different symmetries through
an integral method: the system in question is a nanorod in proximity
of a metallic semi-infinite plane, both of them characterized by a
realistic finite dielectric function.

This paper is organized as follows: in Sec. \ref{subsec:plasmon_modes_interface}
we determine the analytical eigenfrequencies of a nanorod in the proximity
of a flat and semi-infinite imperfect metal \textcolor{black}{using
the bipolar system of coordinates. }Sec. \ref{sec:The-Green-function}
is dedicated to the derivation of the full electrostatic Green's function
(actually a tensor) for the rod plus the metallic half-space. In Sec.
\ref{sec:Determination-of-the} the determination of the fields follows
from the integral method of Eyges, and as a byproduct, the polarizability
tensor is determined. In Sec. \ref{sec:Conclusions} we give our conclusions
and outlook.

\begin{figure}[h]
\centering{}%
\fbox{\begin{minipage}[t]{0.7\columnwidth}%
\begin{center}
\includegraphics[scale=0.9]{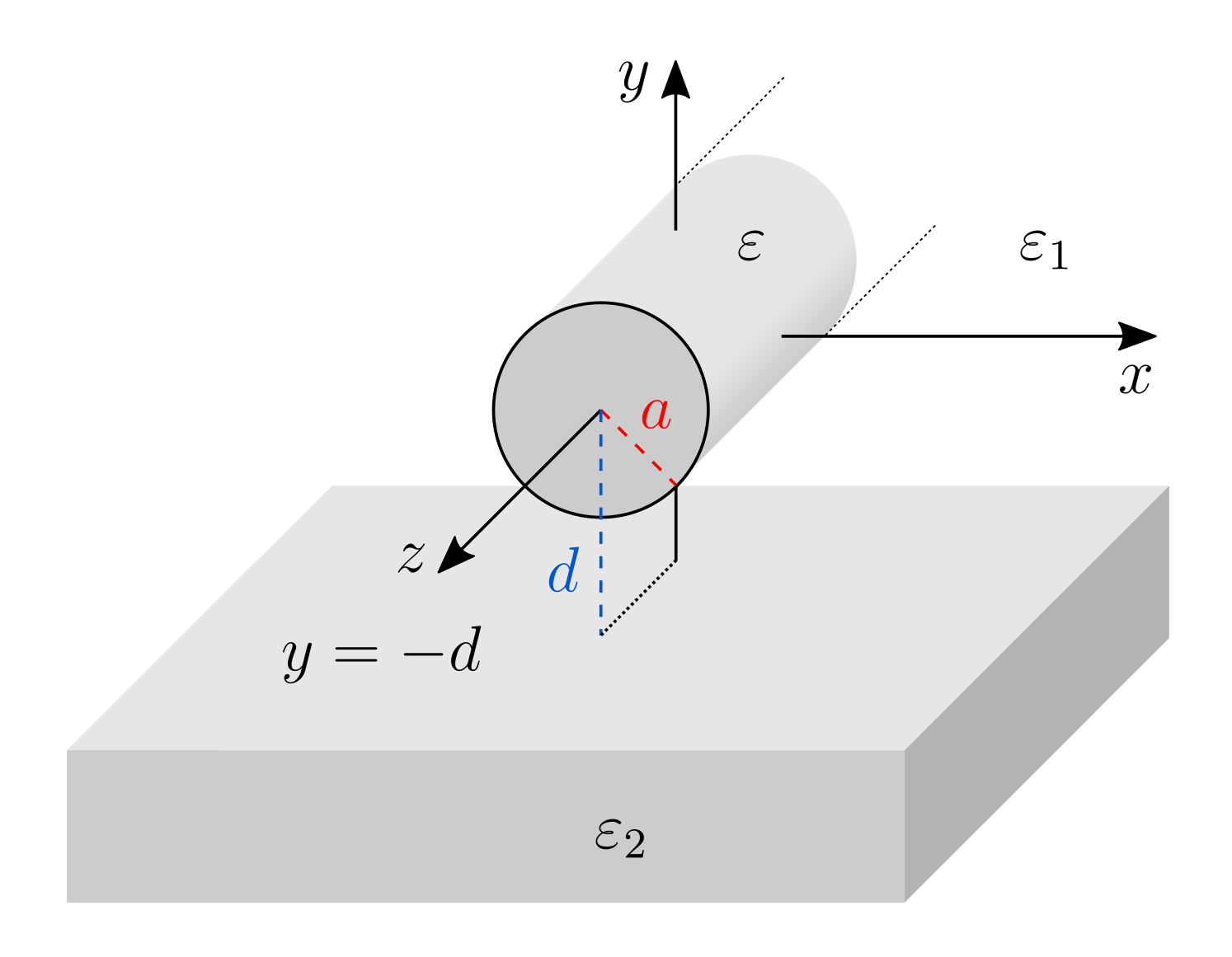}
\par\end{center}
\vspace{-6mm}

\hspace{-12mm}%
\noindent\begin{minipage}[t]{1\columnwidth}%
\begin{center}
\caption{Cylinder of radius $a$ with longitudinal axle coincident with $z$-axis
near an interface between two media, characterized by $\varepsilon_{1}$
and $\varepsilon_{2}$, located at $y=-d$. External potential creates
a uniform electric field in the $xy$-plane.\label{fig:cylinder_near_interface}}
\par\end{center}%
\end{minipage}%
\end{minipage}}
\end{figure}

\section{Plasmon modes in cylinder near an infinite flat interface\label{subsec:plasmon_modes_interface}}

\textcolor{black}{Consider the following electrostatic problem: given
an external electric field, originated from a known electrostatic
potential, acting on an infinitely long cylindrical nanorod in the
vicinity of a dielectric/metallic interface (Fig.~\ref{fig:cylinder_near_interface}),
the aim is to find the total potential everywhere in space. This problem
can be reduced to a two-dimensional one in the $xy$-plane, as the
structure and fields are considered invariant under translation along
the $z$-axis. }

The induced field between the cylinder and the interface can be much
stronger than the external one if the surface plasmon modes are efficiently
excited. These oscillation modes are concentrated near the interface
between two different media and their excitation frequency is related
to the classical plasma frequency $\omega_{{\rm p}}$ according to
the geometry of the system. Solving the Laplace's equation with the
appropriate boundary conditions (BC) allows one to determine the desired
excitation frequencies.

\subsection{Eigenfrequencies: the electrostatic limit case}

\begin{figure}[H]
\centering{}%
\noindent\fbox{\begin{minipage}[t]{1\columnwidth - 2\fboxsep - 2\fboxrule}%
\begin{center}
\includegraphics[scale=0.9]{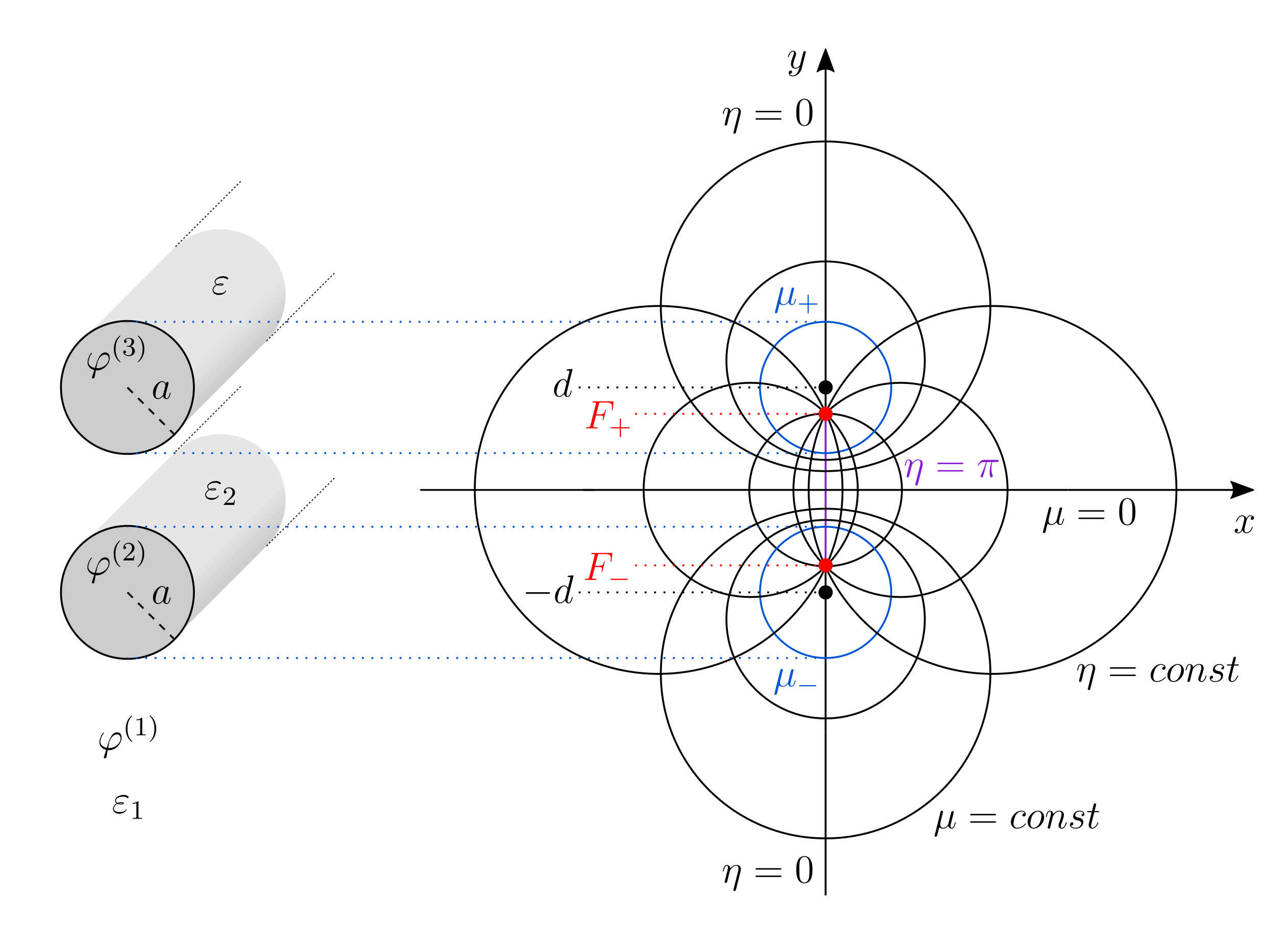}
\par\end{center}
\vspace{-6mm}

\hspace{-12mm}%
\noindent\begin{minipage}[t]{1\columnwidth}%
\caption{System of two cylinders of equal radii in bipolar coordinates with
$F_{-}$ located at $(-u,0)$ and $F_{+}$ located at $(u,0)$. Isosurfaces
$\mu$ and $\eta$ represent orthogonal sets of circles with different
radii. \label{fig:bipolar}}
\end{minipage}%
\end{minipage}}
\end{figure}

Consider first a system of two infinitely long cylinders of radius
$a$, as it is shown by Fig. \ref{fig:bipolar}. For this geometry
the use of bipolar coordinates ($\mu,\eta$) is very well suited \cite{dhondt},
and as one will see later, it allows a simple transition to the configuration
from Fig.~\ref{fig:cylinder_near_interface}. If in the Cartesian
coordinates the foci $F_{-}$ and $F_{+}$ are respectively located
at ($0,-u$) and ($0,u$), the two systems will be related by:
\begin{eqnarray}
x=\frac{u\sin\eta}{\cosh\mu-\cos\eta}, & \,\,\,\,\,\, & y=\frac{u\sinh\mu}{\cosh\mu-\cos\eta},\label{eq:bipolar}
\end{eqnarray}

\noindent where $-\infty<\mu<+\infty$ and $0\leq\eta<2\pi$. The
diagonal metric tensor components are equal, thus leaving the Cartesian
form of the Laplace's equation for the electric potentials unchanged:
\begin{eqnarray}
 &  & \nabla^{2}\varphi^{\left(i\right)}=\frac{\left(\cosh\mu-\cos\eta\right)^{2}}{u^{2}}\left(\frac{\partial^{2}}{\partial\mu^{2}}+\frac{\partial^{2}}{\partial\eta^{2}}\right)\varphi^{\left(i\right)}=0,\,\,\,\,\,\,i=1,2,3.\label{eq:laplace_bipolar}
\end{eqnarray}

\noindent The isosurfaces $-\mu_{-}$ and $\mu_{+}$ coincide with
the circular boundaries of the cylinders characterized by $\varepsilon_{2}$
and $\varepsilon$ respectively, so the following BC are satisfied:

\noindent \numparts
\begin{eqnarray}
 &  & \left.\varphi^{\left(1\right)}\right|_{-\mu_{-}}=\left.\varphi^{\left(2\right)}\right|_{-\mu_{-}},\label{eq:bca}\\
 &  & \left.\varphi^{\left(1\right)}\right|_{\mu_{+}}=\left.\varphi^{\left(3\right)}\right|_{\mu_{+}},\label{eq:bcb}\\
 &  & \varepsilon_{1}\left.\frac{\partial\varphi^{\left(1\right)}}{\partial\mu}\right|_{-\mu_{-}}=\varepsilon_{2}\left.\frac{\partial\varphi^{\left(2\right)}}{\partial\mu}\right|_{-\mu_{-}},\label{eq:bcc}\\
 &  & \varepsilon_{1}\left.\frac{\partial\varphi^{\left(1\right)}}{\partial\mu}\right|_{\mu_{+}}=\varepsilon\left.\frac{\partial\varphi^{\left(3\right)}}{\partial\mu}\right|_{\mu_{+}},\label{eq:bcd}
\end{eqnarray}

\endnumparts

\noindent where the last two equations are written after cancellation
of the metric $h_{\mu}$-coefficients.

~

\noindent The appropriate product solutions of the Laplace's equations
for the three regions in space can be written as follows:

\noindent \numparts
\begin{eqnarray}
 & \varphi_{n}^{\left(1\right)}\left(\mu,\eta\right) & =\left(A_{n}e^{-n\mu}+B_{n}e^{n\mu}\right)\left\{ \begin{array}{c}
\cos\left(n\eta\right)\\
\sin\left(n\eta\right)
\end{array}\right.,\label{eq:2cyl_lap1}\\
 & \varphi_{n}^{\left(2\right)}\left(\mu,\eta\right) & =C_{n}e^{n\mu}\left\{ \begin{array}{c}
\cos\left(n\eta\right)\\
\sin\left(n\eta\right)
\end{array}\right.,\label{eq:2cyl_lap2}\\
 & \varphi_{n}^{\left(3\right)}\left(\mu,\eta\right) & =D_{n}e^{-n\mu}\left\{ \begin{array}{c}
\cos\left(n\eta\right)\\
\sin\left(n\eta\right)
\end{array}\right..\label{eq:2cyl_lap3}
\end{eqnarray}

\noindent \endnumparts

\noindent These solutions satisfy the BC given by Eqs. (\ref{eq:bca})-(\ref{eq:bcd})
only for special values $\varepsilon_{n}$ of $\varepsilon$, so that
one can obtain the subsequent equations for $A_{n}$ and $B_{n}$:

\noindent \numparts
\begin{eqnarray}
 &  & \left(\varepsilon_{1}+\varepsilon_{2}\right)e^{n\mu_{-}}A_{n}-\left(\varepsilon_{1}-\varepsilon_{2}\right)e^{-n\mu_{-}}B_{n}=0,\label{eq:2cyl_1}\\
 &  & \left(\varepsilon_{1}-\varepsilon_{n}\right)e^{-n\mu_{+}}A_{n}-\left(\varepsilon_{1}+\varepsilon_{n}\right)e^{n\mu_{+}}B_{n}=0.\label{eq:2cyl_2}
\end{eqnarray}

\noindent \endnumparts

\noindent This set of homogeneous equations has nonzero solutions
only if the determinant is null. This gives rise to the following
result:
\begin{eqnarray}
 &  & \varepsilon_{n}=-\varepsilon_{1}\frac{\varepsilon_{1}\sinh\left[n\left(\mu_{-}+\mu_{+}\right)\right]+\varepsilon_{2}\cosh\left[n\left(\mu_{-}+\mu_{+}\right)\right]}{\varepsilon_{1}\cosh\left[n\left(\mu_{-}+\mu_{+}\right)\right]+\varepsilon_{2}\sinh\left[n\left(\mu_{-}+\mu_{+}\right)\right]},\label{eq:2cyl_final_result}
\end{eqnarray}

\noindent consisting of a generalization for plasmon modes in two
adjacent circular cross section nanorods derived by Mayergoyz \cite{Mayergoyz2012}.

Now consider that the cylinder of permittivity $\varepsilon$ is located
near a flat interface between media $\varepsilon_{1}$ and $\varepsilon_{2}$.
The solution for this configuration will be a particular case of the
general result from Eq. (\ref{eq:2cyl_final_result}). Transforming
the cylinder with permittivity $\varepsilon_{2}$ into a medium with
a flat surface by setting $\mu_{-}=0$ and renaming $\mu_{+}\rightarrow\mu_{c}$
gives:
\begin{eqnarray}
 &  & \varepsilon_{n}=-\varepsilon_{1}\frac{\left(\varepsilon_{1}+\varepsilon_{2}\right)e^{2n\mu_{c}}+\left(\varepsilon_{2}-\varepsilon_{1}\right)}{\left(\varepsilon_{1}+\varepsilon_{2}\right)e^{2n\mu_{c}}-\left(\varepsilon_{2}-\varepsilon_{1}\right)}.\label{eq:cy_int_final_result}
\end{eqnarray}

\noindent One can relate the parameter $\mu_{c}$ to the cylinder's
radius $a$ and the distance $d$ from its center to the interface
through the following formula:
\begin{eqnarray}
 &  & \mu_{c}={\rm arccosh}\left(d/a\right).\label{eq:mu_da}
\end{eqnarray}

\noindent Characterizing the cylinder by the dielectric function according
to Drude model:
\begin{eqnarray}
 &  & \varepsilon\left(\omega\right)=\varepsilon_{\infty}-\frac{\omega_{{\rm p}}^{2}}{\omega^{2}},\label{eq:dr_nanorod}
\end{eqnarray}

\noindent where $\varepsilon\left(\omega\right)$ is the frequency-dependent
relative permittivity, $\varepsilon_{\infty}$ is the high-frequency
dielectric constant and $\omega_{{\rm p}}$ is the plasma frequency,
one obtains:
\begin{eqnarray}
 &  & \omega_{n}=\omega_{{\rm p}}\left[{\displaystyle \varepsilon_{\infty}+\varepsilon_{1}\frac{\left(\varepsilon_{1}+\varepsilon_{2}\right)e^{2n\mu_{c}}+\left(\varepsilon_{2}-\varepsilon_{1}\right)}{\left(\varepsilon_{1}+\varepsilon_{2}\right)e^{2n\mu_{c}}-\left(\varepsilon_{2}-\varepsilon_{1}\right)}}\right]^{-\frac{1}{2}}.\label{eq:omega_nanorod}
\end{eqnarray}

\noindent When the interface is absent, i.e. $\varepsilon_{1}=\varepsilon_{2}$,
Eq. (\ref{eq:omega_nanorod}) reduces to:
\begin{eqnarray}
 &  & \omega_{n}=\omega_{{\rm p}}\left({\displaystyle \varepsilon_{\infty}+\varepsilon_{1}}\right)^{-\frac{1}{2}}.\label{eq:plasmon_modes}
\end{eqnarray}
Now consider that the cylinder and the lower half-space are made of
the same material, so that $\varepsilon_{n}=\varepsilon_{2}$. For
this special case one obtains a quadratic equation for $\varepsilon_{n}$
in place of Eq. (\ref{eq:cy_int_final_result}), which admits the
following solutions:
\begin{eqnarray}
 &  & \varepsilon_{n}^{\pm}=-\varepsilon_{1}\frac{e^{n\mu_{c}}\pm1}{e^{n\mu_{c}}\mp1}.\label{eq:epsilon_nanorod=00003Dsubstrate}
\end{eqnarray}

\noindent The frequencies able to excite these modes can be written
in terms of hyperbolic functions, thus reading:

\noindent \numparts
\begin{eqnarray}
 &  & \omega_{n}^{+}=\omega_{{\rm p}}\left[{\displaystyle \varepsilon_{\infty}+\varepsilon_{1}\coth\left(\frac{n\mu_{c}}{2}\right)}\right]^{-\frac{1}{2}}\label{eq:omega_n+}\\
 &  & \omega_{n}^{-}=\omega_{{\rm p}}\left[\varepsilon_{\infty}+\varepsilon_{1}\tanh\left(\frac{n\mu_{c}}{2}\right)\right]^{-\frac{1}{2}}.
\end{eqnarray}

\noindent \endnumparts

\noindent As $\mu_{c}\geq0$, when the cylinder is far from the interface
$\left(d\gg a\right)$, these relations also converge to Eq. (\ref{eq:plasmon_modes}).
These oscillation modes correspond to the surface plasmons in free-space
cylinders of circular cross section, as well as for configurations
of infinite flat structures \cite{Mayergoyz2012}. 

\section{The Green's function for the two-dimensional Poisson's equation\label{sec:The-Green-function}}

\subsection{Electrostatic Green's function in two dimensions}

In order to find a general solution for the electrostatic Poisson's
equation with specific BC, we will use the Green's function method.
Consider a potential $\varphi\left(\mathbf{r}\right)$ that satisfies
the Poisson's equation in two dimensions in a medium characterized
by the relative permittivity $\varepsilon$:
\begin{eqnarray}
 &  & \nabla^{2}\varphi\left(\mathbf{r}\right)=-\frac{\sigma\left(\mathbf{r}\right)}{\varepsilon_{0}\varepsilon},\label{eq:poisson_2D}
\end{eqnarray}

\noindent where $\sigma\left(\mathbf{r}\right)$ is the position-dependent
surface charge density and $\varepsilon_{0}$ is the vacuum permittivity.
On the other hand, the potential at $\mathbf{r}=\left(x,y\right)$
due to a point source at $\mathbf{r}'=\left(x',y'\right)$ is given
by the Green's function (GF) $G\left(\mathbf{r}-\mathbf{r}'\right)$,
so that:
\begin{eqnarray}
 &  & \nabla^{2}G\left(\mathbf{r}-\mathbf{r}'\right)=-\frac{\delta\left(\mathbf{r}-\mathbf{r}'\right)}{\varepsilon}.\label{eq:gf_poisson}
\end{eqnarray}
Note that in the expression above only the dimensionless relative
permittivity $\varepsilon$ enters the equation. Using Green's theorem
it can be shown that $\varphi\left(\mathbf{r}\right)$ may be calculated
through the knowledge of the source charge distribution and the GF:
\begin{eqnarray}
 &  & \varphi\left(\mathbf{r}\right)=\frac{1}{\varepsilon_{0}}\int ds'\sigma\left(\mathbf{r}'\right)G\left(\mathbf{r}-\mathbf{r}'\right).\label{eq:fundamental_gf_eq}
\end{eqnarray}
The electrostatic GF in two dimensions for a medium characterized
by the relative permittivity $\varepsilon$ has the form:
\begin{eqnarray}
 &  & G\left(\mathbf{r}-\mathbf{r}'\right)=-\frac{1}{2\pi\varepsilon}\ln\left|\mathbf{r}-\mathbf{r}'\right|,\label{eq:gf_poisson_2d}
\end{eqnarray}
a result that can be obtained by solving explicitly the two-dimensional
Poisson's equation. Another way of deriving this result is through
the Fourier transform. Applying a partial Fourier transform along
the $x$-direction to the Eq. (\ref{eq:gf_poisson}) leads to:
\begin{eqnarray}
 &  & \frac{\partial^{2}}{\partial y^{2}}G\left(q_{x},y\right)-q_{x}^{2}G\left(q_{x},y\right)=-\frac{\delta\left(y\right)}{\varepsilon}.\label{eq:FT_poisson_eq}
\end{eqnarray}

\noindent This form of writing the GF assumes the possibility that
$y$-direction is not translationally invariant. Inserting the Fourier
transforms in the $y$-coordinate in Eq. (\ref{eq:FT_poisson_eq})
and performing the inverse transform after this gives:
\begin{eqnarray}
 &  & G(q_{x},y)=\int_{-\infty}^{+\infty}\frac{dq}{2\pi\varepsilon}\frac{e^{iqy}}{\left(q-iq_{x}\right)\left(q+iq_{x}\right)}=G\left(-q_{x},y\right)=\frac{e^{-\left|q_{x}\right|\left|y\right|}}{2\varepsilon\left|q_{x}\right|}.\label{eq:z_inverse_FT}
\end{eqnarray}
Making now the inverse transform in the $q_{x}$-coordinate, one obtains:
\begin{equation}
G\left(x,y\right)=\int_{-\infty}^{+\infty}\frac{dq_{x}}{2\pi}\cos\left(q_{x}x\right)\frac{e^{-\left|q_{x}\right|\left|y\right|}}{2\varepsilon\left|q_{x}\right|},\label{eq:inverse_FT_qx}
\end{equation}

\noindent where the imaginary part vanished due to the limits of integration.
The integral from Eq. (\ref{eq:inverse_FT_qx}) is formally divergent
due to the logarithm behaviour at the origin. To compute it, we take
its derivative in order to $y$ and assume for now that $y>0$. It
is possible to show that:
\begin{eqnarray}
 &  & \frac{\partial}{\partial y}G\left(x,y\right)=-\frac{1}{2\pi\varepsilon}\frac{y}{x^{2}+y^{2}}\land y>0.\label{eq:inverse_FT_derivative}
\end{eqnarray}

\noindent Making the integration of the previous result over $dy$,
we obtain the desired GF:
\begin{eqnarray}
 &  & G\left(x,y\right)=-\frac{1}{2\pi\varepsilon}\ln\sqrt{x^{2}+y^{2}}+C,\label{eq:G(x,y)}
\end{eqnarray}

\noindent where the constant $C$ (presumably infinite) can be dropped.
For negative $y$ one can repeat the recipe and obtain the same result,
which is equivalent to Eq. (\ref{eq:gf_poisson_2d}). 

In the case of a stratified medium, considering two half-spaces, the
GF in an infinite space with dielectric function $\varepsilon_{j}$
will be:
\begin{eqnarray}
 &  & G_{jj}^{\infty}\left(q_{x},y\right)=\frac{e^{-\left|q_{x}\right|\left|y\right|}}{2\varepsilon_{j}\left|q_{x}\right|}.\label{eq:Gjj}
\end{eqnarray}

\noindent In this case there will be four GFs, depending on which
half-space the source and the observation point are in. If the source
is in the $j=1$ half-space and the observation point in the same
half-space or the one correspondent to $j=2$, the two GFs will respectively
read:

\noindent \numparts
\begin{eqnarray}
 &  & G_{11}\left(q_{x},y,y'\right)=\frac{e^{-\left|q_{x}\right|\left|y-y'\right|}}{2\varepsilon_{1}\left|q_{x}\right|}+Ae^{-\left|q_{x}\right|y},\label{eq:G11}\\
 &  & G_{12}\left(q_{x},y,y'\right)=Be^{\left|q_{x}\right|y}.\label{eq:G12}
\end{eqnarray}

\endnumparts

\noindent Applying the BC so that the source is at $j=1$ ($y'>0$),
we obtain the following equations:

\noindent \numparts
\begin{eqnarray}
 &  & G_{11}\left(q_{x},0,y'\right)=G_{12}\left(q_{x},0,y'\right),\label{eq:bc_green1}\\
 &  & \varepsilon_{1}\frac{\partial}{\partial y}\left.G_{11}\left(q_{x},y,y'\right)\right|_{y=0}=\varepsilon_{2}\frac{\partial}{\partial y}\left.G_{12}\left(q_{x},y,y'\right)\right|_{y=0}.\label{eq:bc_green2}
\end{eqnarray}

\endnumparts

\noindent Equations (\ref{eq:bc_green1}) and (\ref{eq:bc_green2})
embody the continuity of both the electrostatic potential and of the
electric displacement vector normal component. Solving the system
of equations implied by the BC and making the inverse Fourier transform,
we finally obtain the total GFs of the structure:

\noindent \numparts
\begin{eqnarray}
 & G_{11}\left(\mathbf{r},\mathbf{r}'\right) & =-\frac{1}{2\pi\varepsilon_{1}}\ln\left|\mathbf{r}-\mathbf{r}'\right|-\frac{1}{2\pi\varepsilon_{1}}\frac{\varepsilon_{1}-\varepsilon_{2}}{\varepsilon_{1}+\varepsilon_{2}}\ln\left|\mathbf{r}-\tilde{\mathbf{r}}\right|=\label{eq:G11_final}\\
 &  & =\mathcal{G}\left(r,\theta,r',\theta'\right)+\frac{\varepsilon_{1}-\varepsilon_{2}}{\varepsilon_{1}+\varepsilon_{2}}\mathcal{G}_{d}\left(r,\theta,r',\theta'\right),\nonumber \\
 & G_{12}\left(\mathbf{r},\mathbf{r}'\right) & =-\frac{1}{2\pi\varepsilon_{1}}\frac{2\varepsilon_{1}}{\varepsilon_{1}+\varepsilon_{2}}\ln\left|\mathbf{r}-\mathbf{r}'\right|,\label{eq:G12_final}
\end{eqnarray}

\endnumparts

\noindent where $\tilde{\mathbf{r}}=\left(x',-y'\right)$. Note that
the 2D electrostatic GF does not vanish at infinity. Using the same
procedure one can compute $G_{22}$ and $G_{21}$, which correspond
to the situation when the source is in the half-space $j=2$. 

If the cylinder is located at the origin and the interface between
the two media lies at $y=-d$ (Fig. \ref{fig:cylinder_near_interface}),
performing the same procedure as before gives rise to $G_{11}\left(\mathbf{r},\mathbf{r}'\right)$
and $G_{12}\left(\mathbf{r},\mathbf{r}'\right)$ with the same form
as given by Eqs. (\ref{eq:G11_final}) and (\ref{eq:G12_final}),
but with $\tilde{\mathbf{r}}=\left(x',-y'-2d\right)$. The absolute
value of the new $\tilde{\mathbf{r}}$ reads:
\begin{eqnarray}
 &  & \left|\tilde{\mathbf{r}}\right|=\sqrt{\left(x'\right)^{2}+\left(y'+2d\right)^{2}}=\sqrt{(r')^{2}+4dr'\sin\theta'+4d^{2}},\label{eq:rtil}
\end{eqnarray}

\noindent where the relation to polar coordinates is $x'=r'\cos\theta'$
and $y'=r'\sin\theta'$. 

\section{Determination of the fields in the composite system: Eyges' method\label{sec:Determination-of-the}}

Now that the desired Green's function is determined, we can find the
electrostatic potential everywhere from the knowledge of what happens
on the surface of the cylinder. The total potential acting on the
rod is the sum of the external potential $\varphi_{ext}\left(\mathbf{r}\right)$
with the one created by the polarization charges, $\varphi_{\sigma}\left(\mathbf{r}\right)$:
\begin{eqnarray}
 &  & \varphi\left(\mathbf{r}\right)=\varphi_{ext}\left(\mathbf{r}\right)+\varphi_{\sigma}\left(\mathbf{r}\right).\label{eq:total_potential}
\end{eqnarray}

\noindent The dipole moment per unit volume is the 3D polarization
$\mathbf{P}\left(\mathbf{r}'\right)$, where $\mathbf{r}'$ is the
position vector right below the contour line limiting the cross section
of the cylinder. The induced charge density per unit area ($\sigma$)
is related to $\mathbf{P}$ as:
\begin{eqnarray}
 &  & \sigma\left(\mathbf{r}'\right)=\mathbf{\hat{n}}'\cdot\mathbf{P}\left(\mathbf{r}'\right),\label{eq:induced_charge_density}
\end{eqnarray}

\noindent where $\mathbf{\hat{n}}'$ is a unit vector perpendicular
to the surface of the cylinder. The potential due to the new charge
density is calculated through Eq. (\ref{eq:fundamental_gf_eq}). The
total field is given by $\mathbf{E}=-\nabla\varphi(\mathbf{r})$ and
the polarization is related to the aforementioned via the electric
susceptibility, defined as $\chi=\varepsilon-1$, according to:
\begin{eqnarray}
 &  & \mathbf{P}\left(\mathbf{r}'\right)=\varepsilon_{0}\chi\mathbf{E}=-\varepsilon_{0}\chi\nabla'\varphi(\mathbf{r}').\label{eq:polarization_def}
\end{eqnarray}

\noindent Using these definitions and Eq. (\ref{eq:fundamental_gf_eq}),
the total potential reads:
\begin{eqnarray}
 &  & \varphi\left(\mathbf{r}\right)=\varphi_{ext}\left(\mathbf{r}\right)-\chi\int_{s}ds\mathbf{\hat{n}}'\cdot\nabla'\varphi(\mathbf{r}')G\left(\mathbf{r}-\mathbf{r}'\right),\label{eq:psi_total_integral_eq}
\end{eqnarray}

\noindent where the integral is over the contour line limiting the
cross section of the cylinder. We can see that the integral equation
(\ref{eq:psi_total_integral_eq}), firstly proposed by Eyges \cite{Eyges1975},
allows to compute the potential everywhere in space when a uniform
body of arbitrary shape is placed in an external electrostatic field.

The first thing to do is to determine the potential on the cylinder's
surface. With the rod in the upper half-space (Fig. \ref{fig:cylinder_near_interface}),
we have to solve the following equation:
\begin{eqnarray}
 &  & \varphi\left(a,\theta\right)=\varphi_{ext}\left(a,\theta\right)-\chi\int_{s}ds\mathbf{\hat{n}}'\cdot\left.\nabla'\varphi(\mathbf{r}')\right|_{r^{\prime}=a}G_{11}\left(a,\theta,a,\theta'\right).\label{eq:eyges}
\end{eqnarray}

\noindent For the potential inside the cylinder, $\varphi_{in}\left(r,\theta\right)$,
one can write the following expansion:
\begin{eqnarray}
 &  & \varphi_{in}\left(r,\theta\right)=\sum_{l=-\infty}^{+\infty}c_{l}\left(\frac{r}{a}\right)^{\left|l\right|}e^{il\theta},\,\,\,\,\,\,r<a.\label{eq:psi_inside_exp}
\end{eqnarray}

\noindent To determine the coefficients $c_{l}$, we write Eq. (\ref{eq:eyges})
for each index $l$ and expand $\varphi_{in}(a,\theta)$ according
to Eq. (\ref{eq:psi_inside_exp}) on both sides. As the number of
coefficients is infinite, the system has to be truncated at some $\left|l\right|=l_{\max}$,
so that there will be $2l_{\max}+1$ equations to solve, each with
the same number of terms on the right-hand side. Multiplying by $e^{-il\theta}$
and integrating over $\theta$ from $0$ to $2\pi$ gives:
\begin{eqnarray}
 & 2\pi c_{l}= & \int_{0}^{2\pi}d\theta\varphi_{ext}\left(a,\theta\right)e^{-il\theta}\nonumber \\
 &  & -\chi\underset{m\neq0}{\sum_{m=-l_{{\rm max}}}^{l_{{\rm max}}}}\left|m\right|c_{m}\int_{0}^{2\pi}\int_{0}^{2\pi}d\theta'd\theta G_{11}\left(a,\theta,a,\theta'\right)e^{-i\left(l\theta-m\theta'\right)}.\label{eq:eyges_lm}
\end{eqnarray}

\noindent According to Eq. (\ref{eq:G11}), for a cylinder in free
space only the $\mathcal{G}$ part of $G_{11}\left(a,\theta,a,\theta'\right)$
enters the system above, so writing the logarithm argument in polar
coordinates, we have to perform the following integration:
\begin{eqnarray}
 &  & I_{\mathcal{G}}=\int_{0}^{2\pi}\int_{0}^{2\pi}d\theta'd\theta\ln\left[a^{2}\left(\cos\theta-\cos\theta'\right)^{2}+a^{2}\left(\sin\theta-\sin\theta'\right)^{2}\right]e^{-i\left(l\theta-m\theta'\right)}.\label{eq:IG}
\end{eqnarray}

\noindent One can show that all the terms of the integral above with
$l\neq m$ vanish. For $l=m\in\left\{ 0,\pm1\right\} $ it can be
computed analytically, \textcolor{black}{while the general solution
for any $l=m$ was obtained by intuition and numerically confirmed,
thus reading: }

\noindent \begin{equation}
\label{eq:IG_solution}
I_{\mathcal{G}}=
\cases{ 
{\displaystyle -\frac{4\pi^{2}}{\left|m\right|}}, & \(l=m\neq0,\) \cr
8\pi^{2}\ln a, & \(l=m=0,\) \cr
0, & \(l\neq m.\) \cr 
}
\end{equation}

\noindent Note that $8\pi^{2}\ln a$ is always multiplied by zero
due to the presence of $\left|m\right|$ in Eq. (\ref{eq:eyges_lm}),
so that only the $\mathcal{G}(a,\theta,a,\theta')$ terms with $l=m\neq0$
contribute for the potentials. The determination of coefficients $c_{l}$
becomes nontrivial when one has to take the interface into account.
For the Green's function interface part, $\mathcal{G}_{d}(a,\theta,a,\theta')$,
the $y$-coordinate of the source is $-y'-2d$, so the integral to
compute reads:
\begin{eqnarray}
\fl & I_{\mathcal{G}_{d}}= & \int_{0}^{2\pi}\int_{0}^{2\pi}d\theta'd\theta\ln\left[a^{2}\left(\cos\theta-\cos\theta'\right)^{2}+a^{2}\left(\sin\theta+\sin\theta'+\frac{2d}{a}\right)^{2}\right]e^{-i\left(l\theta-m\theta'\right)}.\label{eq:I_Gd}
\end{eqnarray}

\noindent \textcolor{black}{Performing numerical integration for different
ratios $d/a$, we were able to find empirically an analytical result
for the previous integral, valid for $d\geq a$:}

\noindent \begin{equation}
\label{eq:IGd_solution}
I_{\mathcal{G}_{d}}=
\cases{ 
{\displaystyle -i^{\left(l-m\right)}\frac{4\pi^{2}}{\left|m\right|}\frac{\left(\left|l\right|+\left|m\right|-1\right)!}{\left|l\right|!\left(\left|m\right|-1\right)!}\left(\frac{a}{2d}\right)^{\left|l\right|+\left|m\right|},} & \(l\cdot m>0,\) \cr
{\displaystyle -i^{l}\frac{4\pi^{2}}{\left|l\right|}\left(\frac{a}{2d}\right)^{\left|l\right|},} & \(m=0,\,\,l\neq0,\) \cr
{\displaystyle -i^{-m}\frac{4\pi^{2}}{\left|m\right|}\left(\frac{a}{2d}\right)^{\left|m\right|},} & \(l=0,\,\,m\neq0,\) \cr
{\displaystyle 8\pi^{2}\ln\left(2d\right),} & \(l=m=0,\) \cr
0, & \(l\cdot m<0.\) \cr
}
\end{equation} 

\noindent The solutions of the integrals $I_{\mathcal{G}}$ and $I_{\mathcal{G}_{d}}$
presented above allow to make several conclusions about the linear
system given by Eq. (\ref{eq:eyges_lm}). The $\mathcal{G}\left(a,\theta,a,\theta'\right)$
part of the GF only contributes for terms with $l=m$, the diagonal
ones. On the other hand, $\mathcal{G}_{d}\left(a,\theta,a,\theta'\right)$
gives rise to diagonal terms, which can be written in terms of Catalan
numbers, and also to non-diagonal ones when $l$ and $m$ are of the
same sign. For each $l$ the term with $m=0$ is multiplied by zero
according to Eq. (\ref{eq:eyges_lm}), what makes the equation for
$c_{0}$ linearly independent from the others. This way one can write
the following expression for $c_{0}$:
\begin{eqnarray}
 &  & c_{0}=-\frac{\chi}{2\varepsilon_{1}}\frac{\varepsilon_{1}-\varepsilon_{2}}{\varepsilon_{1}+\varepsilon_{2}}\underset{m\neq0}{\sum_{m=-l_{{\rm max}}}^{l_{{\rm max}}}}i^{-m}\left(\frac{a}{2d}\right)^{\left|m\right|}c_{m}.\label{eq:c0}
\end{eqnarray}
For all the other $c$-coefficients the set given by Eq. (\ref{eq:eyges_lm})
can be written in matrix form as:
\begin{eqnarray}
\left[\begin{array}{c}
\vdots\\
c_{-2}\\
c_{-1}\\
c_{1}\\
c_{2}\\
\vdots
\end{array}\right] & = & \left[\begin{array}{c}
\vdots\\
0\\
I_{-1}^{ext}\\
I_{1}^{ext}\\
0\\
\vdots
\end{array}\right]+\left[\begin{array}{cccccc}
\ddots & \vdots & \vdots & \vdots & \vdots & \iddots\\
\cdots & \mathbb{G}_{-2-2} & \mathbb{G}_{-2-1} & 0 & 0 & \cdots\\
\cdots & \mathbb{G}_{-1-2} & \mathbb{G}_{-1-1} & 0 & 0 & \cdots\\
\cdots & 0 & 0 & \mathbb{G}_{11} & \mathbb{G}_{12} & \cdots\\
\cdots & 0 & 0 & \mathbb{G}_{21} & \mathbb{G}_{22} & \cdots\\
\iddots & \vdots & \vdots & \vdots & \vdots & \ddots
\end{array}\right]\left[\begin{array}{c}
\vdots\\
c_{-2}\\
c_{-1}\\
c_{1}\\
c_{2}\\
\vdots
\end{array}\right],\label{eq:eyges_matrix}
\end{eqnarray}

\noindent where the non-null matrix elements $\mathbb{G}_{lm}$ read:
\begin{eqnarray}
 &  & \mathbb{G}_{lm}=-{\displaystyle \frac{\chi}{2\varepsilon_{1}}\left[\delta_{lm}+i^{\left(l-m\right)}\frac{\varepsilon_{1}-\varepsilon_{2}}{\varepsilon_{1}+\varepsilon_{2}}\frac{\left(\left|l\right|+\left|m\right|-1\right)!}{\left|l\right|!\left(\left|m\right|-1\right)!}\left(\frac{a}{2d}\right)^{\left|l\right|+\left|m\right|}\right]},\,\,\,\,\,\,l\cdot m>0,\label{eq:Glm_general}
\end{eqnarray}

\noindent while the non-homogeneous terms are given by:
\begin{eqnarray}
 &  & I_{l}^{ext}=\frac{1}{2\pi}\int_{0}^{2\pi}d\theta\varphi_{ext}\left(a,\theta\right)e^{-il\theta},\,\,\,\,\,\,l=\pm1.\label{eq:nhm_integral}
\end{eqnarray}

\noindent The matrix from Eq. (\ref{eq:eyges_matrix}) can be divided
into two blocks, making its solution simpler. After obtaining the
$c$-coefficients, the potential inside can be written through Eq.
(\ref{eq:psi_inside_exp}), taking $c_{0}$ into account. The potential
outside the cylinder can be determined through the following expression:
\begin{eqnarray}
 &  & \varphi_{out}\left(\mathbf{r}\right)=\varphi_{ext}\left(\mathbf{r}\right)-\chi\underset{m\neq0}{\sum_{m=-l_{{\rm max}}}^{l_{{\rm max}}}}\left|m\right|c_{m}\int_{0}^{2\pi}d\theta'e^{im\theta'}G\left(r,\theta,a,\theta'\right),\label{eq:potential_outside_everywhere}
\end{eqnarray}

\noindent where for the case of configuration from Fig. \ref{fig:cylinder_near_interface}
the GF will be one of the following:

\noindent \numparts
\begin{eqnarray}
\fl & G_{11}\left(r,\theta,a,\theta'\right)= & -\frac{1}{4\pi\varepsilon_{1}}\ln\left[\left(r\cos\theta-a\cos\theta'\right)^{2}+\left(r\sin\theta-a\sin\theta'\right)^{2}\right]\nonumber \\
\fl &  & -\frac{1}{4\pi\varepsilon_{1}}\frac{\varepsilon_{1}-\varepsilon_{2}}{\varepsilon_{1}+\varepsilon_{2}}\ln\left[\left(r\cos\theta-a\cos\theta'\right)^{2}+\left(r\sin\theta+a\sin\theta'+2d\right)^{2}\right],\label{eq:G11(r)}\\
\fl & G_{12}\left(r,\theta,a,\theta'\right)= & -\frac{1}{4\pi\varepsilon_{1}}\frac{2\varepsilon_{1}}{\varepsilon_{1}+\varepsilon_{2}}\ln\left[\left(r\cos\theta-a\cos\theta'\right)^{2}+\left(r\sin\theta-a\sin\theta'\right)^{2}\right].\label{eq:G12(r)}
\end{eqnarray}

\noindent \endnumparts

\noindent We note here that the terms
\begin{eqnarray}
{\rm r}=\frac{\varepsilon_{1}-\varepsilon_{2}}{\varepsilon_{1}+\varepsilon_{2}}, & \,\,\,\,\,\, & {\rm t}=\frac{2\varepsilon_{1}}{\varepsilon_{1}+\varepsilon_{2}}\label{eq:r_t}
\end{eqnarray}

\noindent are respectively the Fresnel reflection and transmission
coefficients of an electromagnetic wave with normal incidence on an
interface in the electrostatic limit. The integration over $\theta'$
in Eq. (\ref{eq:potential_outside_everywhere}) with $G_{11}\left(r,\theta,a,\theta'\right)$
will determine the potential above the interface, while using $G_{12}\left(r,\theta,a,\theta'\right)$
one will obtain what happens to the potential in the substrate. 

Solving the problem with $l_{{\rm max}}=1$ for $d=2a$ and $\varphi_{ext}=-E_{0}r\sin\theta$,
by fixing $\theta=\pi/2$ one is able to write the following result:

\noindent \begin{equation}
\label{eq:potential_at_x=0}
\fl
\varphi\left(r,\frac{\pi}{2}\right)= \cases{
{\displaystyle \varphi_{in}=i\left(\pm\frac{2r}{a}+\frac{\chi}{4\varepsilon_{1}}\frac{\varepsilon_{1}-\varepsilon_{2}}{\varepsilon_{1}+\varepsilon_{2}}\right)c_{1},} & \(r<a,\) \cr
{\displaystyle {\displaystyle \varphi_{out}=-i\frac{\chi}{\varepsilon_{1}}\left(\pm\frac{a}{r}-\frac{\varepsilon_{1}-\varepsilon_{2}}{\varepsilon_{1}+\varepsilon_{2}}\frac{1}{4\pm r/a}\right)c_{1}\mp E_{0}r},} & \(r>a,\,\,\,y>0,\) \cr
{\displaystyle {\displaystyle \varphi_{out}}=-i\frac{\chi}{\varepsilon_{1}}\frac{2\varepsilon_{1}}{\varepsilon_{1}+\varepsilon_{2}}\frac{a}{r}c_{1}+E_{0}r,} & \(y<-d,\) \cr } \end{equation}

\noindent where the sign of $r$ in $\varphi_{in}$ and $\varphi_{out}$
is positive if $y>0$ and negative for $y<0$, while the coefficient
$c_{1}$ reads:
\begin{eqnarray}
 &  & c_{1}=\frac{iE_{0}a}{2+{\displaystyle \frac{\chi}{\varepsilon_{1}}\left(1+\frac{1}{16}\frac{\varepsilon_{1}-\varepsilon_{2}}{\varepsilon_{1}+\varepsilon_{2}}\right)}}.\label{eq:c1}
\end{eqnarray}

\noindent 
\begin{figure}[H]
\noindent\fbox{\begin{minipage}[t]{1\columnwidth - 2\fboxsep - 2\fboxrule}%
\begin{flushleft}
\begin{tabular*}{2cm}{@{\extracolsep{\fill}}>{\centering}m{8cm}>{\centering}m{8cm}}
\begin{centering}
\vspace{3.2mm}
\par\end{centering}
\includegraphics[scale=0.75]{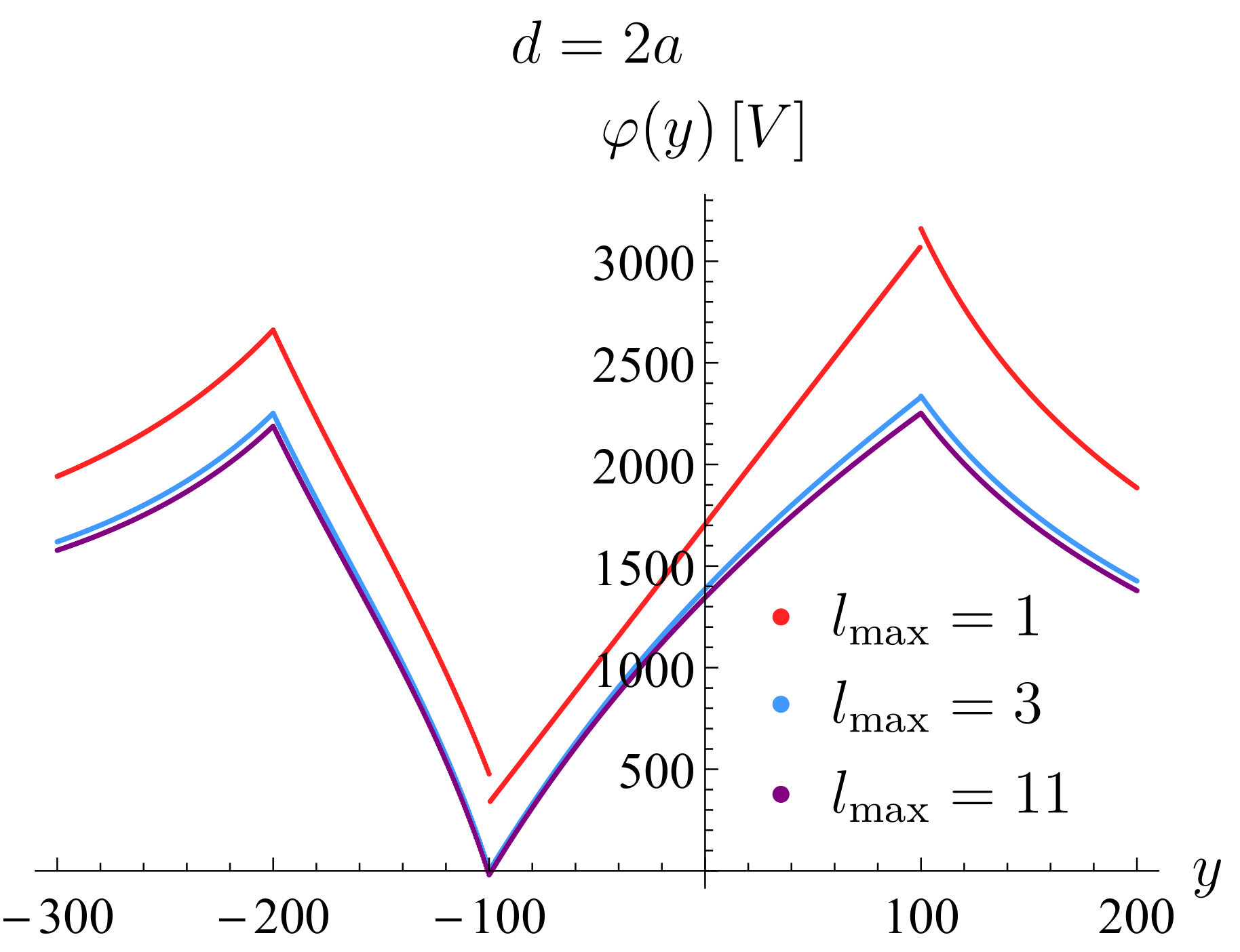}
\begin{centering}
\vspace{5mm}
\par\end{centering}
\centering{}\hspace{-70mm}\textbf{\small{}A} & \includegraphics[scale=0.24]{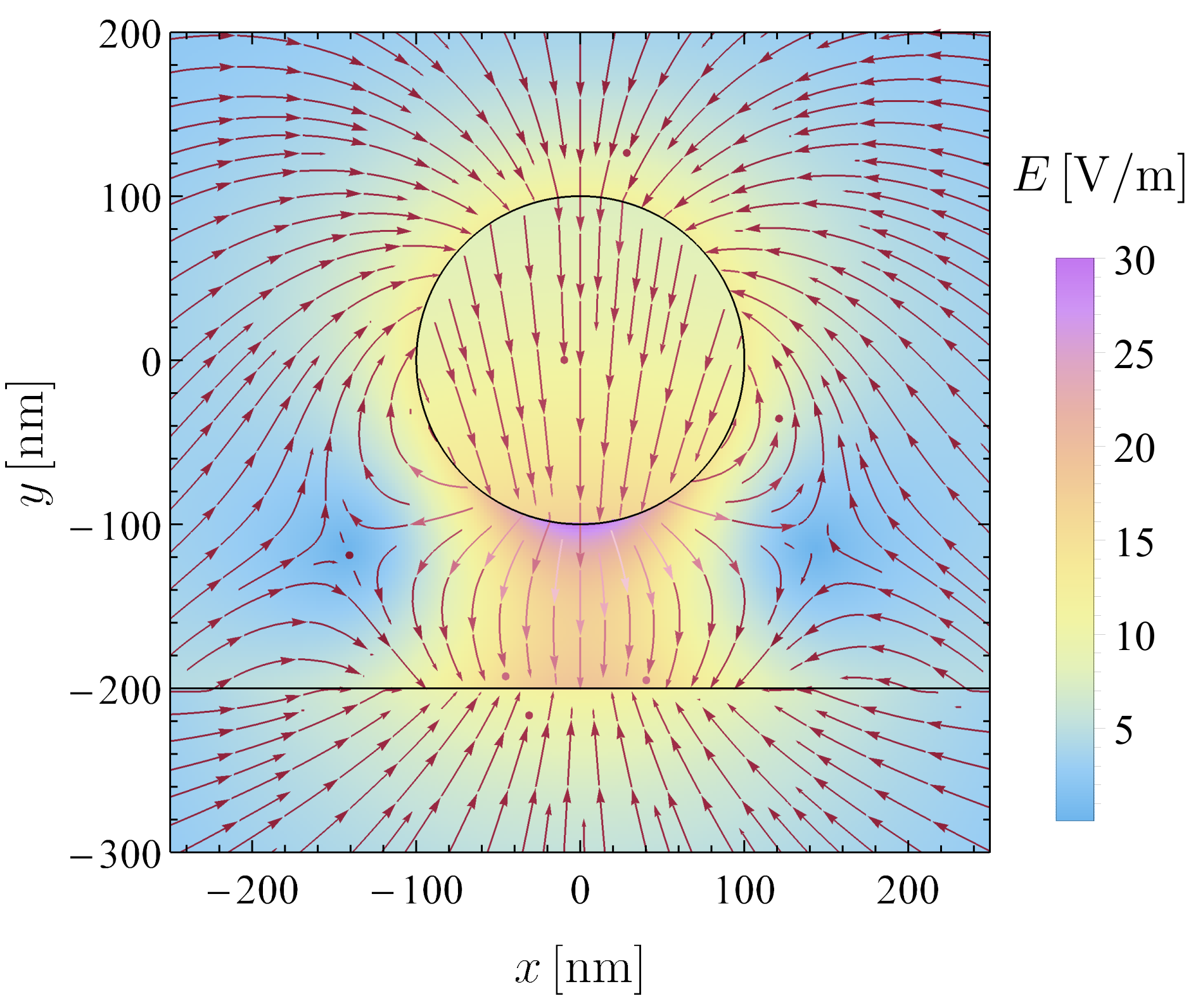}
\begin{centering}
\vspace{-11mm}
\par\end{centering}
\centering{}\hspace{-70mm}\textbf{\small{}B}\tabularnewline
 & \tabularnewline
\includegraphics[scale=0.24]{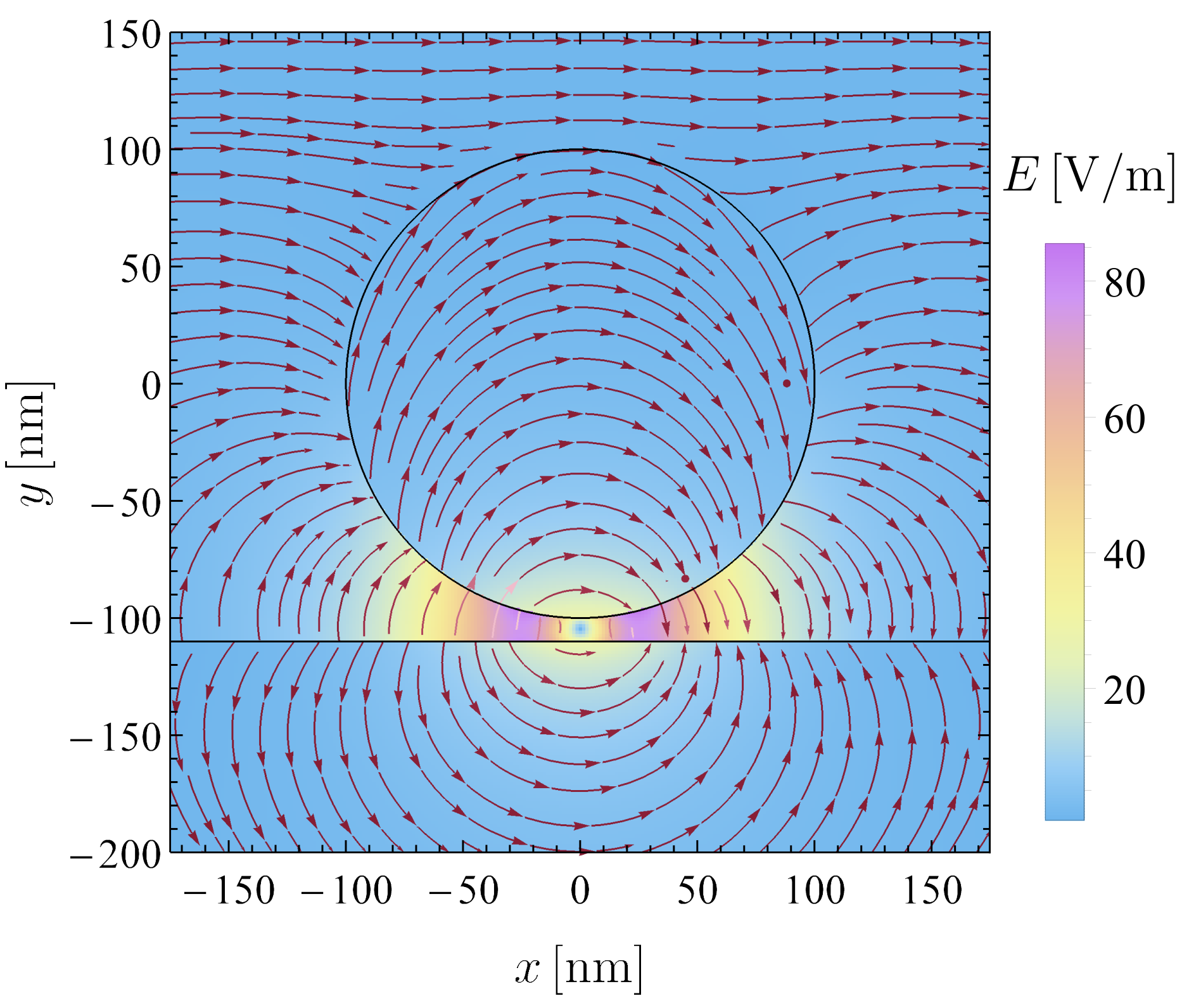}

\vspace{-6mm}
\centering{}\hspace{-70mm}\textbf{\small{}C} & \includegraphics[scale=0.24]{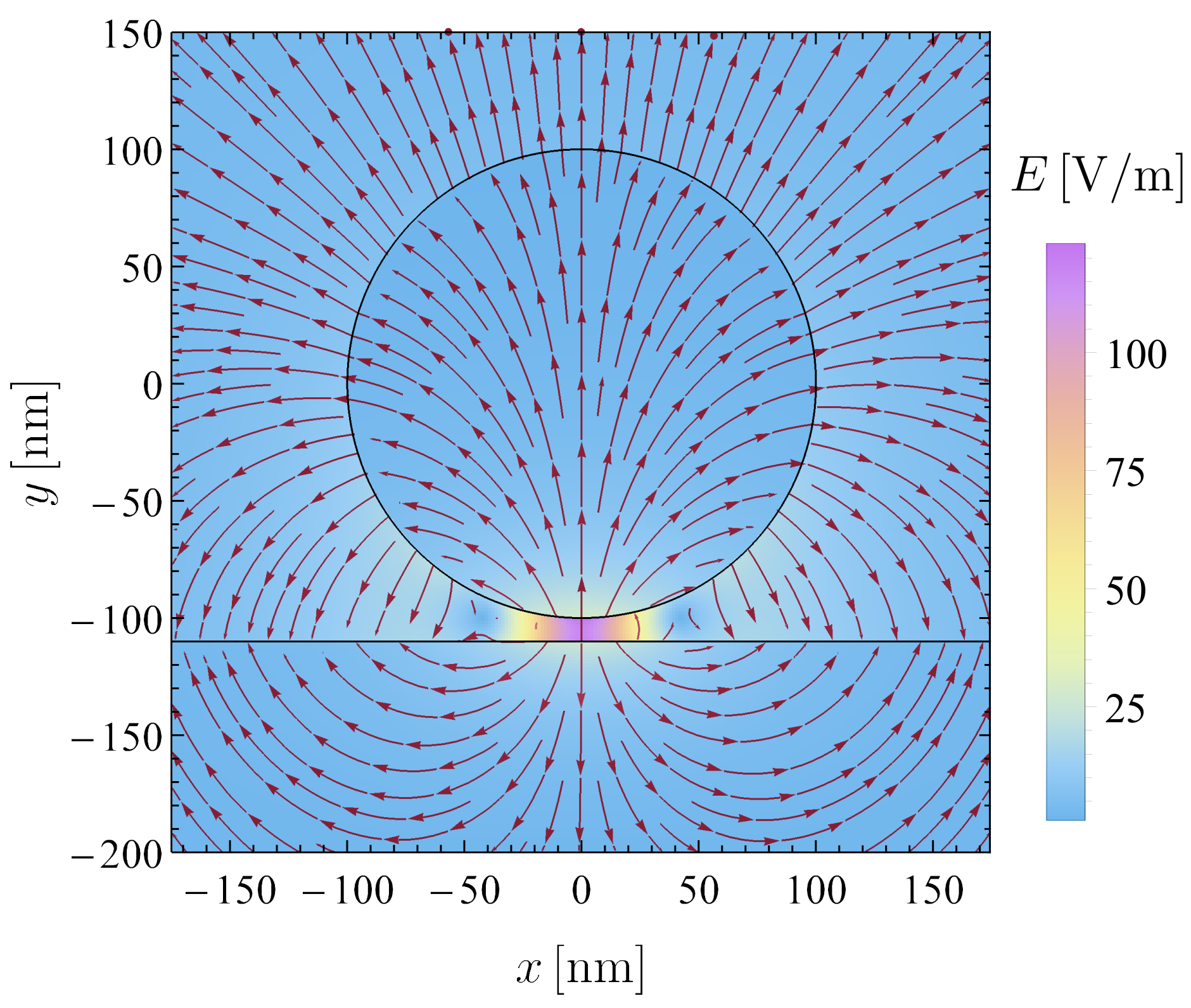}

\vspace{-6mm}
\centering{}\hspace{-70mm}\textbf{\small{}D}\tabularnewline
\end{tabular*}
\par\end{flushleft}
\vspace{-6mm}

\hspace{-12mm}%
\noindent\begin{minipage}[t]{1\columnwidth}%
\caption{Ag nanorod of radius $a=100\,{\rm nm}$ near a vacuum-Ag interface
with $\varepsilon_{1}=1$ and $\varepsilon_{2}=\varepsilon_{{\rm Ag}}\left(\omega\right)$
according to Drude model ($\varepsilon_{\infty}=5$, $\hbar\omega_{{\rm p}}=9.04\,{\rm eV}$,
$\Gamma=21.25\,{\rm meV}$ \cite{Zeman1987}) subjected to an external
electric field of intensity $E_{0}=1\,{\rm V/m}$. \textbf{A}: Total
electric potential along $y$-axis for $\mathbf{E}_{ext}=E_{0}\hat{\mathbf{y}}$
and $d=2a$ ($l_{{\rm max}}=1,3,11$).\textbf{ B}: Total electric
field for $\mathbf{E}_{ext}=E_{0}\hat{\mathbf{y}}$ and $d=2a$ ($l_{\max}=11$).
\textbf{C}: Total electric field for $\mathbf{E}_{ext}=E_{0}\hat{\mathbf{x}}$
and $d=1.1a$ ($l_{\max}=15$). \textbf{D}: Total electric field for
$\mathbf{E}_{ext}=E_{0}\hat{\mathbf{y}}$ and $d=1.1a$ ($l_{\max}=15$).\label{fig:total_electric_field_near_interface}}
\end{minipage}%
\end{minipage}}
\end{figure}

The $c$-coefficients and hence the potential inside the cylinder
can be obtained analytically. The general result for the potential
outside the cylinder can be calculated numerically, allowing to determine
the electric field everywhere through $\mathbf{E}=-\nabla\varphi_{out}\left(\mathbf{r}\right)$.
Plotting the total potential along the $y$-axis for the case of an
Ag nanorod at a distance $d=2a$ from a vacuum-Ag interface (Fig.
\ref{fig:total_electric_field_near_interface}A), it is possible to
see that with $l_{{\rm max}}=1$ the potential is slightly discontinuous
across the nanorod's surface, but converges very quickly to a continuous
solution, and at $l_{{\rm max}}=3$ it almost overlaps the solution
with $l_{{\rm max}}=11$. When the cylinder is closer to the surface,
the convergence is slower and higher $l_{{\rm max}}$ is needed to
solve the problem correctly. Different field patterns are obtained
depending on the distance $d$ and also on the direction of $\mathbf{E}_{ext}$.
Inside the nanorod the field lines follow the direction of the external
electric field for $l_{\max}=1$, but become progressively distorted
by the scattered potential as one takes more higher-order terms into
account. When $\mathbf{E}_{ext}$ is perpendicular to the media border
(Figs. \ref{fig:total_electric_field_near_interface}B and \ref{fig:total_electric_field_near_interface}D),
the maximal field magnitude occurs exactly in the middle of the gap
between the rod and the interface. In this region the stream lines
converge (Fig. \ref{fig:total_electric_field_near_interface}B) or
diverge (Fig. \ref{fig:total_electric_field_near_interface}D) from
the flat surface, depending on the distance the cylinder is located
from the substrate. This corresponds respectively to accumulation
of negative or positive charge on the area just below the cylinder.
One can also see that the closer the nanorod is to the substrate,
the higher is the magnitude of the field in the gap between the two.
If $\mathbf{E}_{ext}$ has parallel incidence (Fig. \ref{fig:total_electric_field_near_interface}C),
then there are two regions where the intensity of the electric field
is higher, each with different sign, having lower magnitude comparatively
to the perpendicular incidence. 

~

\subsection{Polarizability of the nanorod}

Using the relation between the polarization $\mathbf{P}$ and the
electric field $\mathbf{E}$ from Eq. (\ref{eq:polarization_def}),
we can write for the dipole moment $\mathbf{p}$:
\begin{eqnarray}
 &  & \mathbf{p}=\int_{V}dV\mathbf{P}=-\varepsilon_{0}\chi\int dV\nabla'\varphi\left(\mathbf{r}'\right),\label{eq:pol1}
\end{eqnarray}

\noindent what can be transformed into a surface integral using the
divergence theorem:
\begin{eqnarray}
 &  & \mathbf{p}=-\varepsilon_{0}\chi\int ds\hat{\mathbf{n}}'\varphi\left(\mathbf{r}'\right).\label{eq:pol2}
\end{eqnarray}

\noindent Expressing the total potential through Eq. (\ref{eq:psi_inside_exp})
and writing $\hat{\mathbf{n}}'=\hat{\mathbf{r}}'=\left(\cos\theta,\sin\theta\right)$
we obtain:
\begin{eqnarray}
 & \mathbf{p} & =-\varepsilon_{0}\chi a\sum_{l=-\infty}^{+\infty}c_{l}\int_{0}^{2\pi}d\theta e^{il\theta}\left(\cos\theta,\sin\theta\right)=\nonumber \\
 &  & =-\varepsilon_{0}\chi a\pi\left[c_{-1}+c_{1},-i\left(c_{-1}-c_{1}\right)\right].\label{eq:pol3}
\end{eqnarray}

\noindent Since the polarizability $\alpha$ satisfies the relation
$\mathbf{p}=\alpha\mathbf{E}_{0}$, having in mind that the $c$-coefficients
contain the factor $aE_{0}$, the diagonal components of the polarizability
tensor read:

\noindent \numparts
\begin{eqnarray}
 & \alpha_{xx} & =-\frac{\varepsilon_{0}\chi a}{E_{0}}\pi\left(c_{-1}+c_{1}\right),\label{eq:pol4}\\
 & \alpha_{yy} & =i\frac{\varepsilon_{0}\chi a}{E_{0}}\pi\left(c_{-1}-c_{1}\right).\label{eq:pol5}
\end{eqnarray}

\endnumparts

\noindent For external field with normal incidence the coefficients
$c_{1}$ and $c_{-1}$ have the same absolute value and opposite sign,
being the first given by Eq. (\ref{eq:c1}) for $l_{{\rm max}}=1$
and $d=2a$, so that only $\alpha_{yy}$ exists. For parallel incidence
the coefficients $c_{\pm1}$ are equal, hence only $\alpha_{xx}$
does not vanish. Note that contrary to a finite three-dimensional
body, the polarizability of the infinite cylinder has units of $\varepsilon_{0}\times{\rm Area}$
instead of $\varepsilon_{0}\times{\rm Volume}$.

\begin{figure}[b]
\noindent\fbox{\begin{minipage}[t]{1\columnwidth - 2\fboxsep - 2\fboxrule}%
\begin{flushleft}
\begin{tabular*}{2cm}{@{\extracolsep{\fill}}>{\centering}m{8cm}>{\centering}m{8cm}}
\begin{centering}
\includegraphics[scale=0.8]{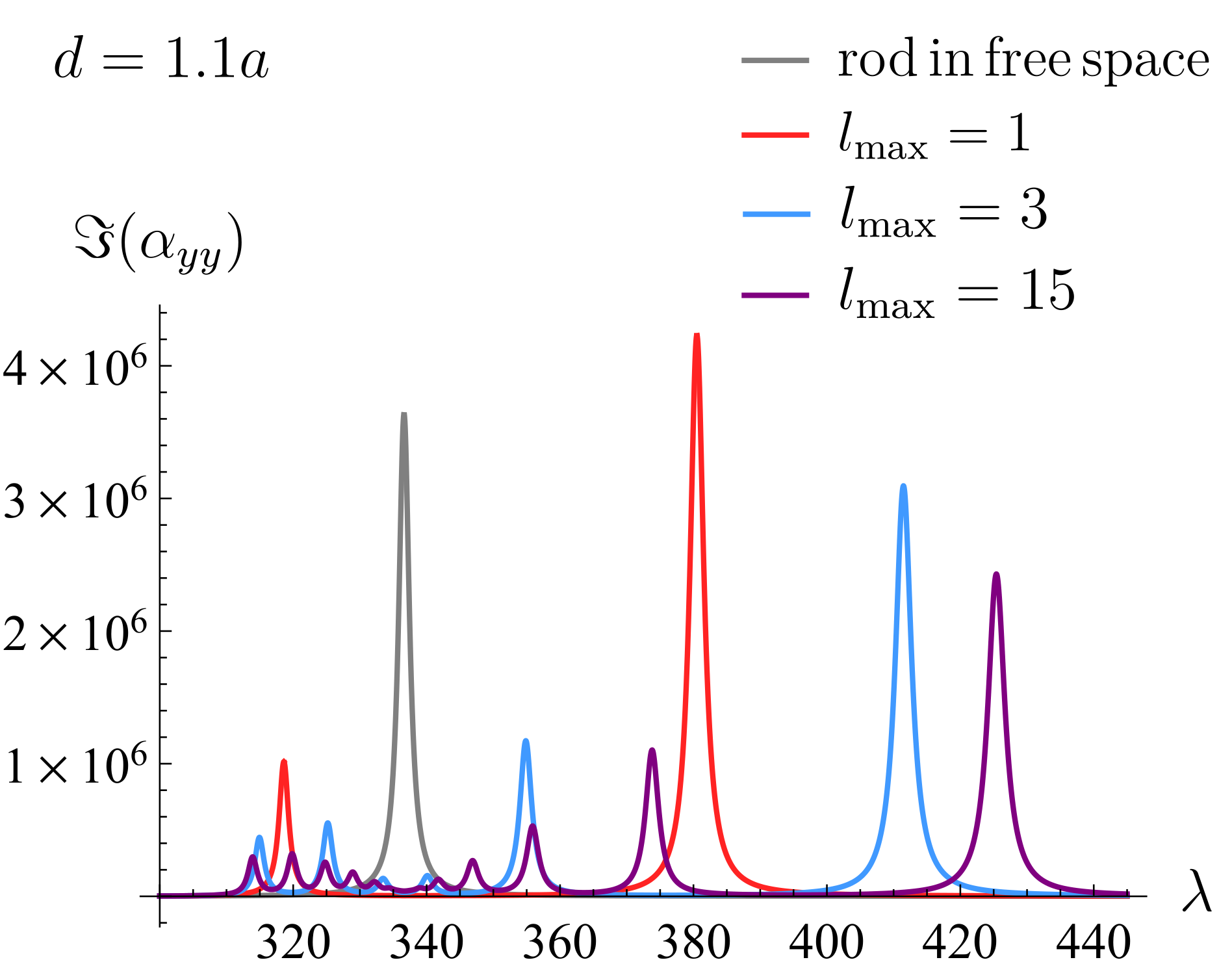}
\par\end{centering}
\begin{centering}
\vspace{-5.75mm}
\par\end{centering}
\centering{}\hspace{-70mm}\textbf{\small{}A} & \begin{centering}
\includegraphics[scale=0.8]{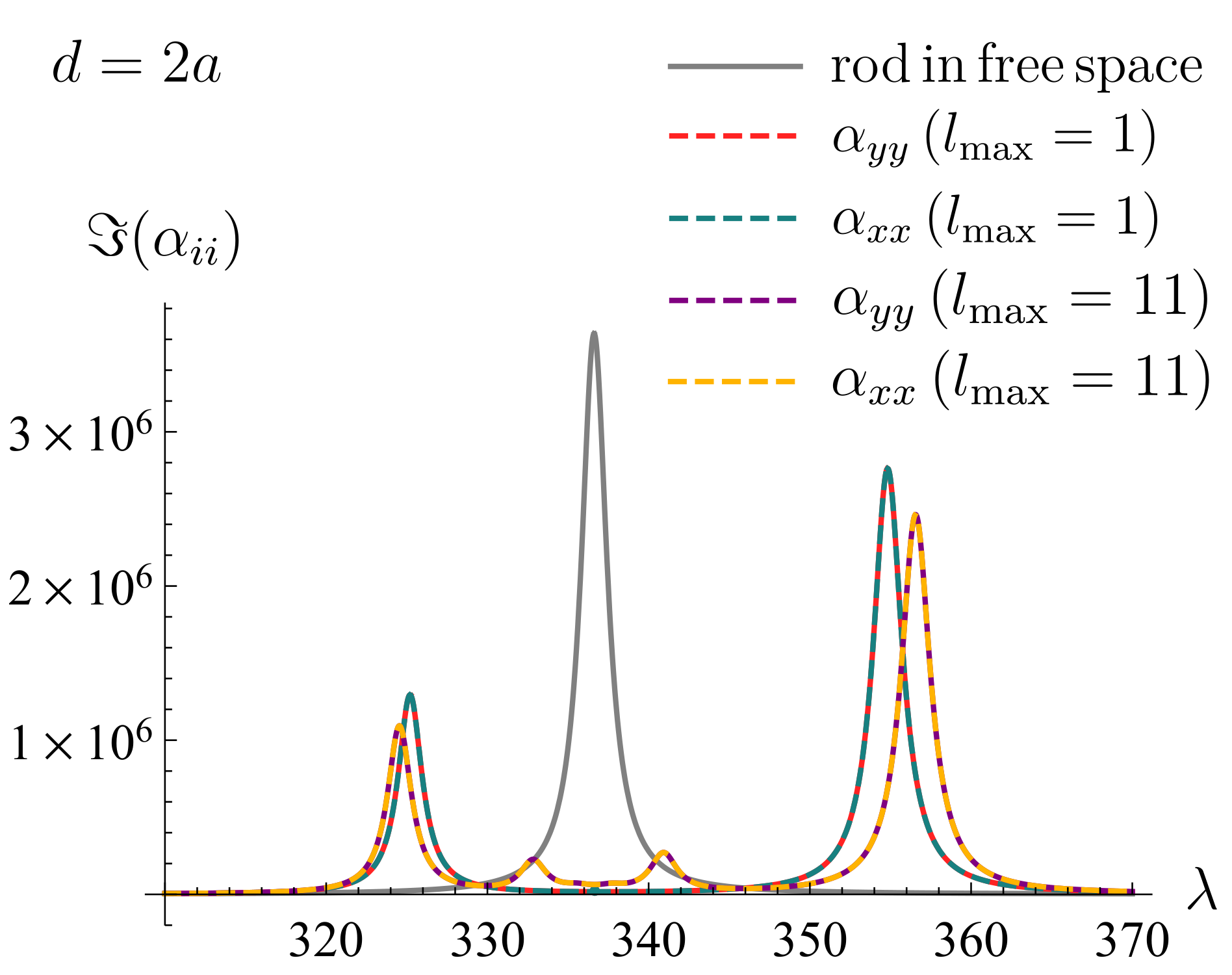}
\par\end{centering}
\begin{centering}
\vspace{-5.75mm}
\par\end{centering}
\centering{}\hspace{-70mm}\textbf{\small{}B}\tabularnewline
\end{tabular*}
\par\end{flushleft}
\vspace{-6mm}

\hspace{-12mm}%
\noindent\begin{minipage}[t]{1\columnwidth}%
\caption{Imaginary part of polarizability tensor components (in units of $\varepsilon_{0}\times[{\rm nm}]^{2}$)
as a function of the wavelength (${\rm nm}$) for an Ag nanorod of
radius $a=100\,{\rm nm}$ near a vacuum-Ag interface with $\varepsilon_{1}=1$
and $\varepsilon_{2}=\varepsilon_{{\rm Ag}}\left(\omega\right)$ according
to Drude model ($\varepsilon_{\infty}=5$, $\hbar\omega_{{\rm p}}=9.04\,{\rm eV}$,
$\Gamma=21.25\,{\rm meV}$ \cite{Zeman1987}). The grey line shows
the imaginary part of the polarizability in the absence of an interface
near the rod and presents a single peak. The introduction of an interface
splits this peak in several. \textbf{A}: Imaginary part of $\alpha_{yy}$
for $d=1.1a$ and $l_{\max}=1,3,15$. \textbf{B}: Imaginary part of
$\alpha_{xx}$ and $\alpha_{yy}$ for $d=2a$ and $l_{\max}=1,11$.
\label{fig:polarizability_near_interface}}
\end{minipage}%
\end{minipage}}
\end{figure}

The polarizability tensor components depend on the distance $d$,
which is also hidden in the $c$-coefficients. Plots of the imaginary
part of the diagonal components of $\alpha_{ij}$ (Fig. \ref{fig:polarizability_near_interface})
show that the smaller the distance $d$, the longer it takes for the
sum of the modes to converge. On Fig. \ref{fig:polarizability_near_interface}B
we can also see that $\alpha_{xx}$ and $\alpha_{yy}$ coincide, expressing
the isotropy of the polarizability. The position of the relative maxima
of $\Im\left(\alpha_{ii}\right)$ is exactly predicted by the solution
of Laplace's equation with appropriate BC performed in Section \ref{subsec:plasmon_modes_interface}.
The wavelengths of the natural modes are given by:
\begin{eqnarray}
 &  & \lambda_{n}^{\pm}=\lambda_{{\rm p}}\sqrt{{\displaystyle \varepsilon_{\infty}+\varepsilon_{1}\frac{e^{n\mu_{c}}\pm1}{e^{n\mu_{c}}\mp1}}},\label{eq:lambda_n}
\end{eqnarray}
where $\lambda_{{\rm p}}$ is the wavelength correspondent to the
classical plasma frequency and $\mu_{c}={\rm arccosh}\left(d/a\right)$.
For $d=1.1a$ the peaks at $313.9\,{\rm nm}$ and $425.4\,{\rm nm}$
correspond respectively to $\lambda_{1}^{-}$ and $\lambda_{1}^{+}$,
the maxima at $319.8\,{\rm nm}$ and $373.8\,{\rm nm}$ respectively
to $\lambda_{2}^{-}$ and $\lambda_{2}^{+}$ and so on. Note that
it is necessary to use larger values of $l_{\max}$ to obtain the
precise position of the maxima for $d=1.1a$ correspondent to $\lambda_{5}^{\pm}$,
for example, as the modes with higher $n$ take longer to converge.
When $n\rightarrow\infty$, the geometrical factor $\left(e^{n\mu_{c}}\pm1\right)/\left(e^{n\mu_{c}}\mp1\right)$
tends to 1, so for any ratio $d/a$ the wavelengths of the peaks will
converge to:
\begin{eqnarray}
 &  & \lambda=\lambda_{{\rm p}}\sqrt{{\displaystyle \varepsilon_{\infty}+\varepsilon_{1}}},\label{eq:lambda_plasmon}
\end{eqnarray}

\noindent what corresponds to the natural frequency of plasmon modes
in cylinders of circular cross section and in infinite flat structures
\cite{Mayergoyz2012}. 

One can use the imaginary part of the polarizability tensor components
to calculate the extinction cross section. The last one is defined
as the sum of the absorption and scattering cross sections and can
be written as:
\begin{eqnarray}
 &  & \sigma_{ex}=\sigma_{ab}+\sigma_{sc}=\frac{k}{\varepsilon_{0}}\Im\left(\alpha_{ii}\right),\label{eq:extinction_cross_section}
\end{eqnarray}

\noindent where $k=\omega/c$. We note here that the extension of
the cross section in two dimensions has units of length and not of
area, as in three-dimensional problems.

\section{Conclusions and outlook\label{sec:Conclusions}}

In this paper we have studied analytically a plasmonic system composed
of a nanorod in the vicinity of a plasmonic metallic surface, forming
a small gap between both structures. The study was performed in the
electrostatic and local approximations. The first approximation is
accurate in the regime $\lambda\gg a$, where $a$ is the rod's radius,
so that retardation plays no significant role. The second approximation
is justifiable as long as the gap between the rod and metallic plane
is not too small. The inclusion of nonlocal effects \cite{Hildebrandt2004}
poses considerably additional difficulties to the calculations because
an extra field component and an additional boundary condition appear
\cite{Nat-Comm_Andre}. However, for distances of the order of a few
nanometers or smaller, nonlocal effects rise in importance and prevent
the evergrowing enhancement of field intensity as the gap between
the nanostructure and the semi-infinite plane is reduced. Additionally,
for very narrow gaps quantum effects start to play an important role
and we enter the realm of quantum plasmonics. Having an analytical
description of both nonlocal and quantum effects is the goal of a
future publication. The inclusion of two-dimensional materials in
the system considered in this paper is an extension worth pursuing
due to the recent demonstration of extraordinary confinement of graphene
acoustic plasmons in metallic nanostructures of the form considered
here \cite{Epstein1219}. The original experiments in the hybrid system
composed of graphene and nanocubes showed extraordinary field enhancement
and confinement. The same is expected if nanorods are deposited on
top of graphene instead of nanocubes. The nanorod, however, represents
considerable interest itself, since the shape of the field will be
different, which may open an additional degree of freedom when it
comes to the application of this type of hybrid system for molecular
sensing in the mid-IR. The extension of this work to include 2D materials,
at the lowest level of approximation, is relatively simple. The main
new complication comes from the new form of one of the boundary conditions,
where the derivatives normal to 2D materials are now discontinuous,\textcolor{black}{{}
due to the charge density accumulated, for example, in graphene. The
integrals in Eyges' method become, however, more complicated. }

\section*{Acknowledgments }

N.M.R.P. thanks André Gonçalves and Bruno Amorim for many discussions
on the topic of plasmonics. N.M.R.P. acknowledges support by the Portuguese
Foundation for Science and Technology (FCT) in the framework of the
Strategic Funding UIDB/04650/2020, support from the European Commission
through the project ``Graphene-Driven Revolutions in ICT and Beyond''
(Ref. No. 881603, CORE 3), COMPETE 2020, PORTUGAL 2020, FEDER and
the FCT through projects POCI-01-0145-FEDER-028114.

\newpage{}

\bibliographystyle{iopart-num}

\providecommand{\newblock}{}

\end{document}